\documentclass{moriond}

\def\Journal#1#2#3#4{{#1} {\bf #2}, #3 (#4)}

\def\be{\begin{equation}}
\def\ee{\end{equation}}
\def\bea{\begin{eqnarray}}
\def\eea{\end{eqnarray}}

\newcommand{\Photo}{}

\usepackage{hepnames}

\newcommand{\Bmunu}{\PB \rightarrow \Pmu \Pnum}

\newcommand{\Blnugamma}{\PB \rightarrow \Plepton \Pneutrino \gamma}
\newcommand{\BmuN}{\PB \rightarrow \Pmu N}

\newcommand{\Vub}{V_\mathrm{ub}}
\newcommand{\BKstarLL}{\PB \rightarrow \PKstar \Pleptonplus\Pleptonminus}

\newcommand{\alphaem}{\alpha_\text{em}}
\newcommand{\GF}{G_\text{F}}

\begin{document}
\vspace*{4cm}
\title{Study of Lepton Universality at Belle }

\author{Markus Tobias Prim for the Belle Collaboration}

\address{Karlsruhe Institute of Technology (KIT), Institute of Experimental Particle Physics (ETP), Wolfgang-Gaede-Str. 1, 76131 Karlsruhe, Germany}

\maketitle\abstracts{
	The Belle Collaboration presented three recent Belle analyses: The search for $\Blnugamma$ ($\Plepton = \Pe, \Pmu$) with improved hadronic tagging, the search for $\Bmunu$ with inclusive tagging and the test of lepton universality in $\BKstarLL$ ($\Plepton = \Pe, \Pmu$) decays.
}

\section{Introduction}
The Belle experiment collected an integrated luminosity of $711\, \text{fb}^{-1}$ at the $\Upsilon(4\text{S})$ resonance, corresponding to $772\times10^6$ $\PB\APB$-events, using the Belle detector and the KEKB asymmetric-energy $\APelectron\Pelectron$ collider. In this manuscript we present three recent results from the Belle Collaboration.

\section{Search for the rare decay $\Blnugamma$ with improved hadronic tagging}
\subsection{Motivation}
\begin{figure}[b]
	\centering
	\includegraphics{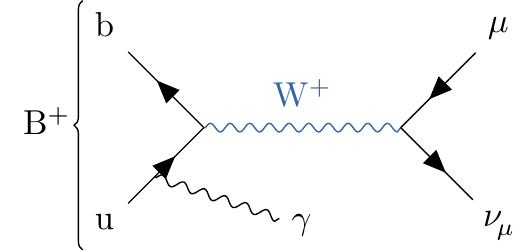}
	\caption{Leading order diagram for the process $\Blnugamma$.}
	\label{app:feynman:Bmunugamma}
\end{figure}

Purely leptonic \PB decays are helicity suppressed in the Standard Model (SM). This helicity suppression can be lifted if a high energetic final state photon is radiated from, e.g.\, the \Pup~quark in the initial bound state. This leads to the excitation of the \PB~meson to a virtual S=1 resonance, whose decay into a $\Pmu\Pnum$-pair is not helicity suppressed. The dominant Feynman diagram to this process is shown in Figure~\ref{app:feynman:Bmunugamma}.
Although the suppression by the electromagnetic coupling constant $\alphaem$ arises in this process, the lifting of the helicity suppression results in an enhanced branching ratio for the process. The differential decay rate for the process as a function of the \Pgamma energy is given by
\begin{equation}
\frac{\mathrm{d}\Gamma(\Blnugamma)}{\mathrm{d}E_{\Pgamma}} = \frac{\alphaem \GF^2 |\Vub|^2}{6 \pi^2} m_{\PB} E_{\Pgamma}^3 \left(1 - \frac{2E_{\Pgamma}}{m_{\PB}}\right) \times \left( \left|F_\text{V}\right|^2 + \left|F_\text{A} + \frac{e_{\Plepton} f_{\PB}}{E_{\Pgamma}}\right|^2 \right)\ ,
\end{equation}
with Fermi's coupling constant $\GF$, the CKM matrix element $\Vub$, the \PB meson mass $m_{\PB}$, the \PB decay constant $f_{\PB}$, and the lepton charge $e_{\Plepton}$.
The axial-vector $F_\text{A}$ and vector $F_\text{V}$ hadronic currents parametrize the hadronic transition. In the limit of high photon energies $E_{\Pgamma} > 1\,\text{GeV}$ the form factors can be expanded\cite{beneke} as
\begin{equation}
F_\text{V/A}(E_{\Pgamma}) = \frac{e_{\Pup}f_{\PB}m_{\PB}}{2E_{\Pgamma} \lambda_{\PB}(\mu)} R(E_{\Pgamma, \mu}) + \xi(E_{\Pgamma}) \pm \Delta \xi(E_{\Pgamma})\ ,
\end{equation}
where $e_{\Pup}$ is the charge of the $\Pup$~quark. Further, the factor $R(E_{\Pgamma, \mu})$ describes the photon emissions from the light spectator quark in the \PB meson. The $\xi$ and $\Delta \xi$ terms contain a symmetry conserving and breaking part for both form factors $F_{V/A}$ and are suppressed by factors of $1/m_{\PB}$ and $1/(2E_{\Pgamma})$.
The parameter $\lambda_{\PB}$ is closely related to the first inverse moment of the leading twist \PB meson light cone distribution amplitude $\phi_+$ in the high energy limit, $\lambda_{\PB}^{-1} = \int_{0}^{\infty} \mathrm{d}w \phi_+(w)$, where $w$ is the light-cone momentum. This parameter is an important input into calculations of QCD factorization, which is used to predict non-leptonic \PB meson decays.\cite{qcd-fac-1,qcd-fac-2,qcd-fac-3} The parameter can not be derived from first principles and is unknown. For $\lambda_{\PB}$ in the order of several hundred MeV and for photon energies $E_{\Pgamma} > 1\, \text{GeV}$ the partial branching ratio $\Delta\mathcal{B}(\Blnugamma)$ is expected to be of $\mathcal{O}(10^{-6})$.\cite{beneke}

\subsection{Results}
\begin{figure}
	\centering
	\includegraphics[width=0.45\linewidth]{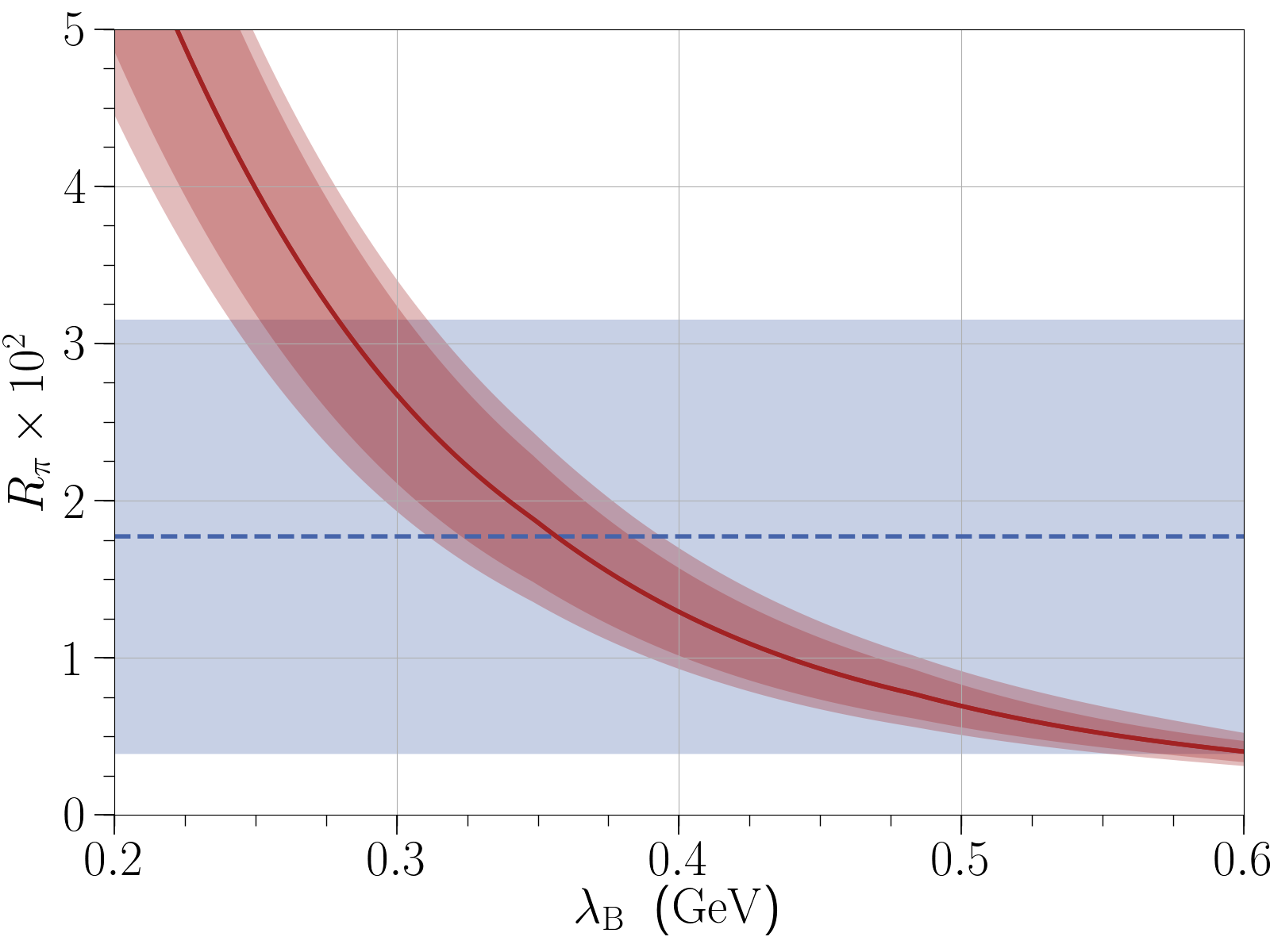}
	\caption{The theory prediction for $R_{\Ppi}$ with the 1 $\sigma$ uncertainty band is shown in red, where the dark red band is the theoretical uncertainty and the light red contains the uncertainty from the light-cone distribution amplitude model dependence. The measured value with its uncertainty is shown in blue.}
	\label{fig:Blnugamma:lambda}
\end{figure}

The key improvements of the presented analysis are an improved \PB tagging algorithm, the Full Event Interpretation\cite{fei} (FEI), improved modeling of the background arising from semileptonic \PB decays, and a new method to determine the upper limit on $\lambda_{\PB}$ via the ratio
\begin{equation}
	R_{\Ppi} = \frac{\Delta \mathcal{B}(\Blnugamma)}{\mathcal{B}(\PB \rightarrow \Ppi \Plepton \Pneutrino)}\ ,
\end{equation}
where the dependence on $|\Vub|$ cancels.
The theory prediction and measured ratio for $R_{\Ppi}$ is shown in Figure~\ref{fig:Blnugamma:lambda}. For a detailed description of the analysis see Reference~\cite{gelb}.
The determined upper limit of the partial branching ratio with $E_{\Pgamma} > 1\, \text{GeV}$ is
\begin{equation}
	\Delta \mathcal{B}(\Blnugamma) < 3.0\times 10^{-6}\ \text{at}\ 90\%\, \text{CL}\ ,
\end{equation}
resulting in an upper limit of
\begin{equation}
	\lambda_{\PB} > 0.24\ \text{GeV}\ .
\end{equation}

\section{Search for $\Bmunu$ and $\BmuN$ with inclusive tagging}
\subsection{Motivation}
\begin{figure}
	\centering
	\includegraphics[width=0.3\linewidth]{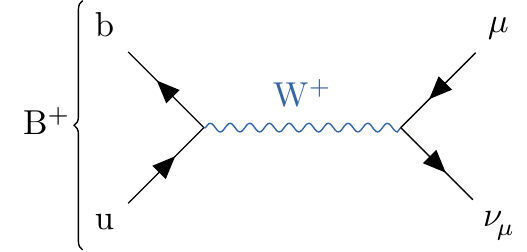}
	\includegraphics[width=0.3\linewidth]{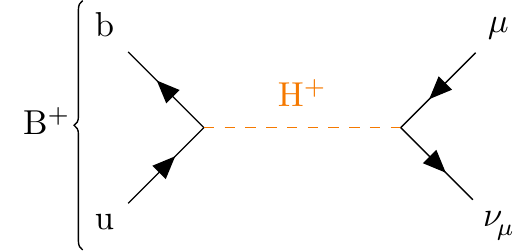}
	\includegraphics[width=0.3\linewidth]{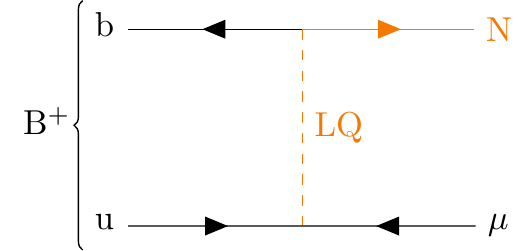}
	\caption{Tree level Feynman diagram for the SM process $\Bmunu$ (left). Tree level Feynman diagram for the 2HDM process $\Bmunu$ (middle). One possible realization for the process $\BmuN$ with Leptoquarks (right).}
	\label{app:feynman:Bmunu}
\end{figure}

The branching ratio of the decay $\Bmunu$ can be calculated within the SM and is predicted to be
\begin{equation}
	\mathcal{B}(\Bmunu)_\text{SM} = \frac{\GF^2 m_{\PB} m_{\Plepton}^2}{8 \pi} \left(1 - \frac{m_{\Plepton}^2}{m_{\PB}^2}\right)^2 f_{\PB}^2 |\Vub|^2 \tau_{\PB}\ .
	\label{eq:Bmunu:BR}
\end{equation}
The tree level Feynman diagram of the process is shown on the left of Figure~\ref{app:feynman:Bmunu}. 
From Equation~\ref{eq:Bmunu:BR} it is apparent that the decay is both helicity and CKM suppressed, which results in the small branching ratio of $\mathcal{B}(\Bmunu)_\text{SM} \approx (4.26 \pm 0.7) \cdot 10^{-7}$. Although rare, the decay has a very clean experimental signature of a high energetic monochromatic lepton momentum.
Models beyond the SM can influence this branching ratio and/or the nominal momentum of the lepton. In the following two models are discussed.

Within the 2 Higgs Doublet Models (2HDM) the SM Higgs sector is extended by an additional Higgs doublet, which results in a total of five Higgs bosons after electroweak symmetry breaking, including a charged Higgs boson which can mediate the decay $\Bmunu$. The leading order Feynman diagram where the charged Higgs mediates the decay is shown in the middle of Figure~\ref{app:feynman:Bmunu}. Depending on the parameters of the 2HDM model, the interference between the SM and 2HDM processes is either constructive or destructive, modifying the experimental signature by enhancing or reducing the observed branching ratio. In this manuscript we investigate the type-II and type-III 2HDM models\cite{type2,type3-1,type3-2}, where the modification of the branching ratio is given by
\begin{equation}
	\mathcal{B}(\Bmunu)_\text{Type-II} = \mathcal{B}(\Bmunu)_\text{SM} \times \left| 1 - \frac{m_{\PB}^2 \tan^2 \beta}{m_{\PHiggsplus}^2} \right|^2\ ,
\end{equation}
with the ratio of the vacuum expectation values of the two Higgs Doublets $\tan\beta$ and the mass of the charged Higgs boson $m_{\PHiggsplus}$, and
\begin{equation}
	\mathcal{B}(\Bmunu)_\text{Type-III} = \mathcal{B}(\Bmunu)_\text{SM} \times \left| 1 + \frac{m_{\PB}^2}{m_{\Pbottom}m_{\Plepton}} \frac{C_\text{R} - C_\text{L}}{C_\text{SM}} \right|^2\ ,
\end{equation}
where the coefficients $\text{C}_\text{R/L}$ encode the new physics contribution.

The search for the decay $\Bmunu$ allows to search for a sterile neutrino in the final state in a model independent way. One possible realization with a Leptoquark as mediator is shown on the right of Figure~\ref{app:feynman:Bmunu}. The only requirement imposed on the sterile neutrino is a lifetime large enough to not decay inside of the detector volume. With a massive sterile neutrino in the final state, the signal peak of the monochromatic lepton momentum shifts to lower energies, thus revealing a second signal contribution in the lepton momentum spectrum, or an enhancement of the measured branching ratio if the two signal peaks are indistinguishable.

\subsection{Results}
\begin{figure}
	\centering
	\includegraphics[width=0.45\linewidth]{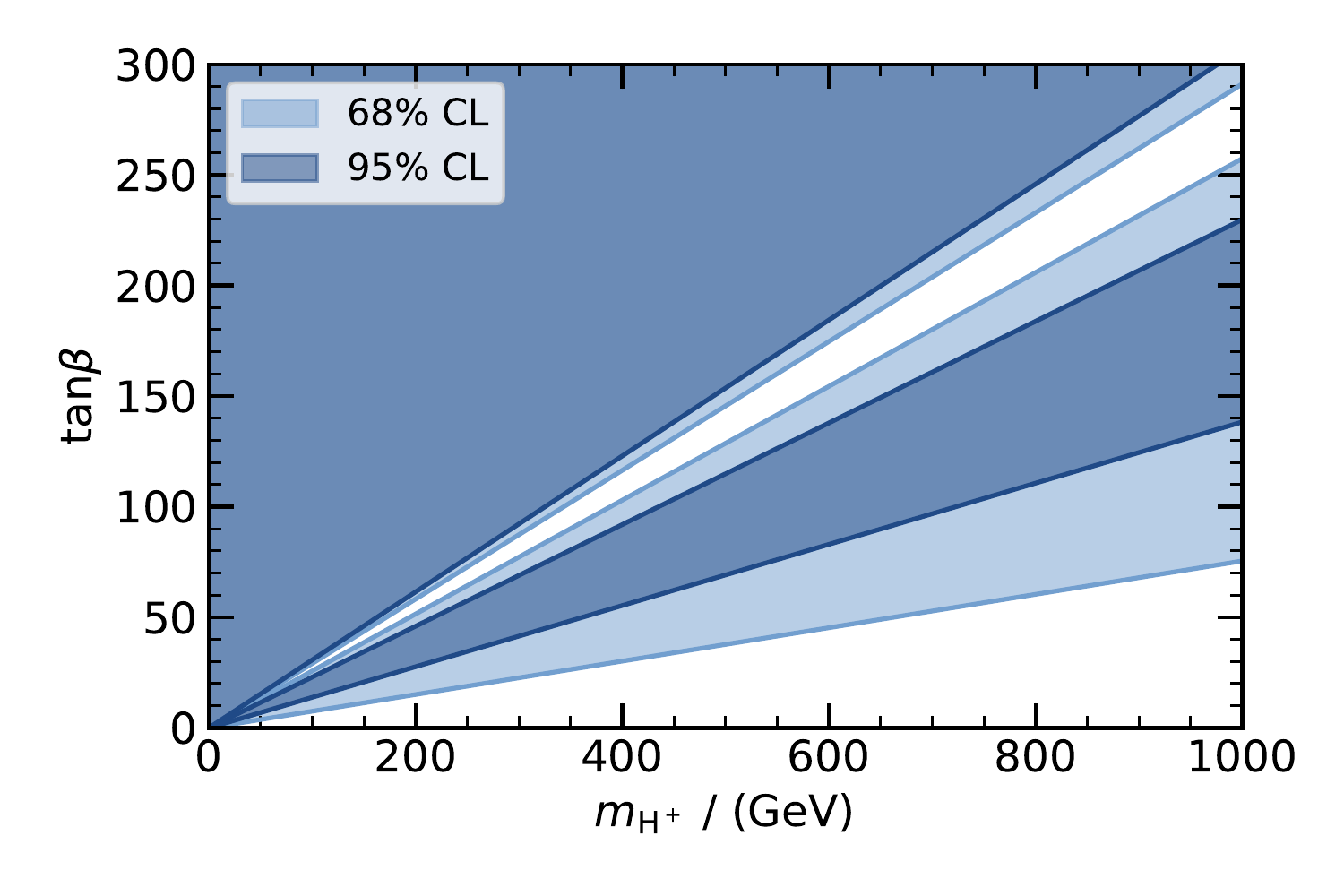}
	\includegraphics[width=0.45\linewidth]{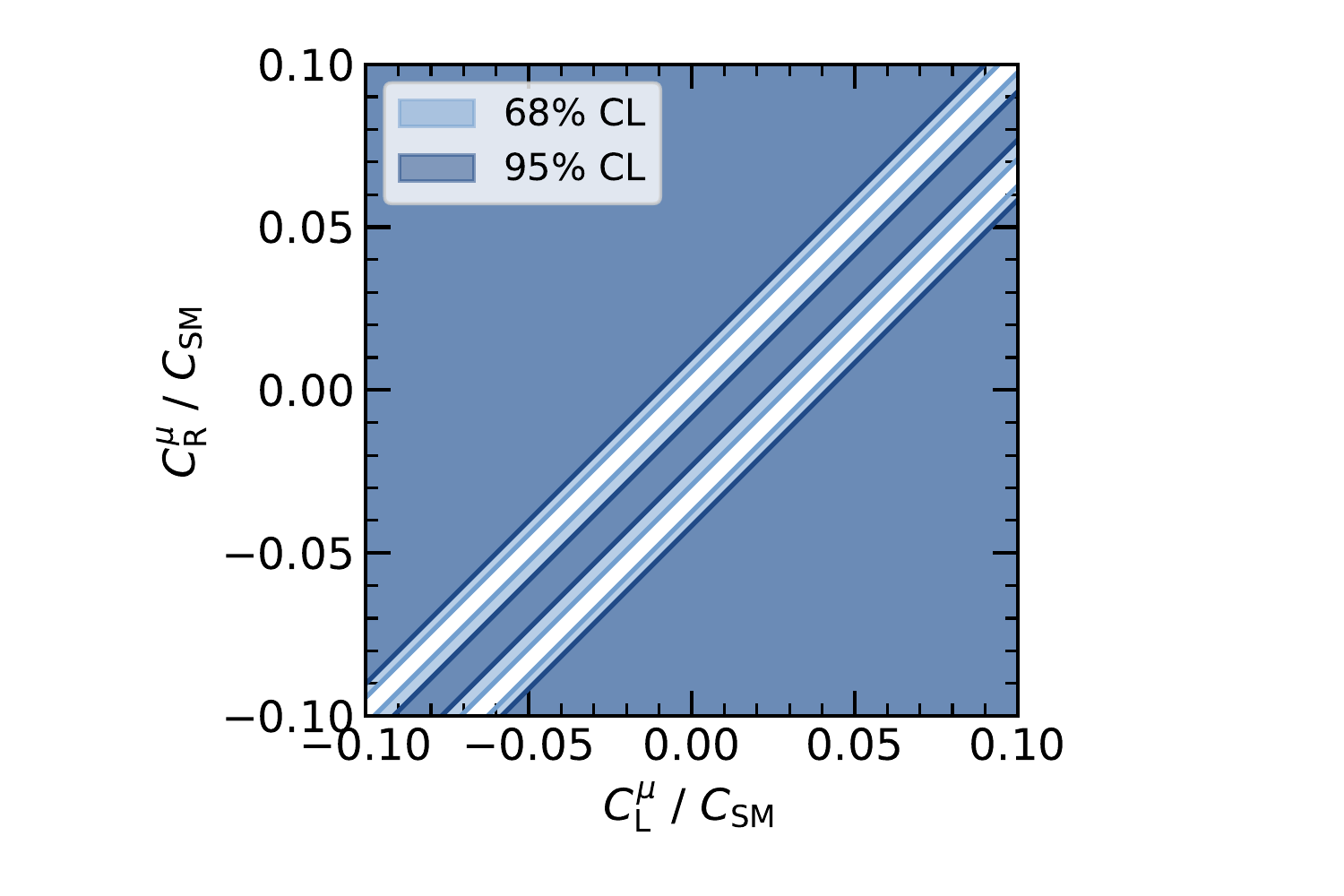}	
	\caption{68\% and 95\% CL excluded model parameter space for the 2HDM type-II ($\tan \beta$, $m_{\PHiggsplus}$) and type-III ($\text{C}_\text{R}$, $\text{C}_\text{L}$) are shown.}
	\label{fig:Bmunu:2hdm}
\end{figure}

\begin{figure}
	\centering
	\includegraphics[width=0.5\linewidth]{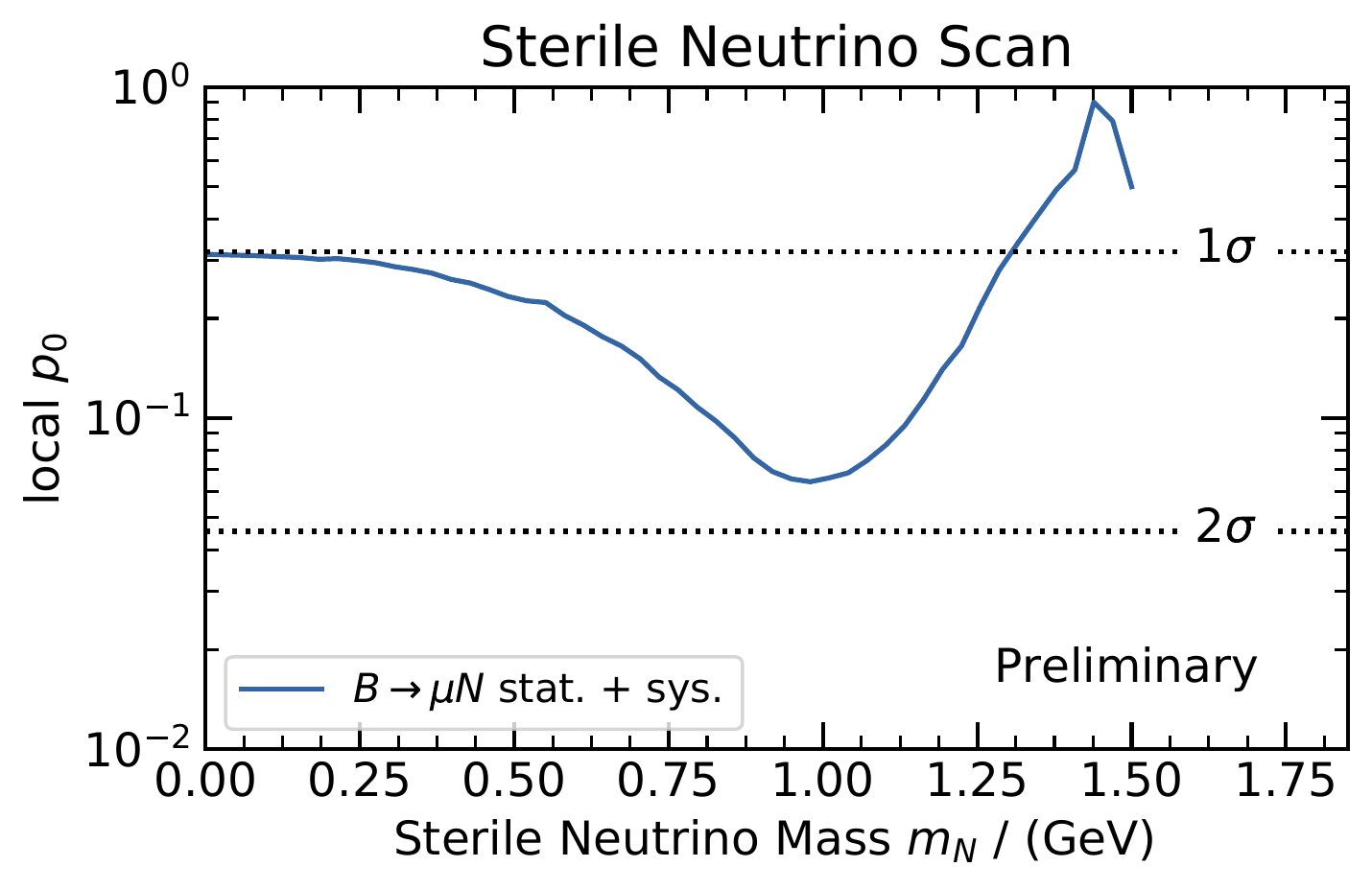}
	\caption{The observed local $p_0$ values for the sterile neutrino search $\BmuN$ are shown with the SM process $\Bmunu$ included. If the SM process is accounted for, no significant excess is observed. The largest deviation from the background-only hypothesis is at $m_N = 1\ \text{GeV}$. No correction for the look elsewhere effect is included.}
	\label{fig:Bmunu:sterile}
\end{figure}

The key improvements of the presented analysis are an inclusive tagging approach with a dedicated calibration to improve the momentum resolution, which allows to measure the monochromatic signal peak in the parent \PB meson, an improved description of the semileptonic $\PB\rightarrow X_{\Pup}\Plepton \Pneutrino$ decays and an improved modeling the continuum background. For a detailed description see the upcoming publication of the analysis.
The measured branching ratio is
\begin{equation}
	\mathcal{B}(\Bmunu) = (5.3 \pm 2.0 \pm 0.9) \times 10^{-7}
\end{equation}
with a significance of 2.8 standard deviations over the background only hypothesis.
The allowed region for the type-II and type-III 2HDM model parameters are calculated with the measured central value of the branching ratio and are shown in Figure~\ref{fig:Bmunu:2hdm}. 
The scan for a sterile neutrino in the measured lepton momentum spectrum was performed by fixing the SM process to its expectation. The local $p_0$ value in the scanned mass range is shown in Figure~\ref{fig:Bmunu:sterile}. No significant excess was observed.

\section{Test of lepton flavor universality in $\BKstarLL$ decays at Belle}

\subsection{Motivation}
\begin{figure}
	\centering
	\includegraphics{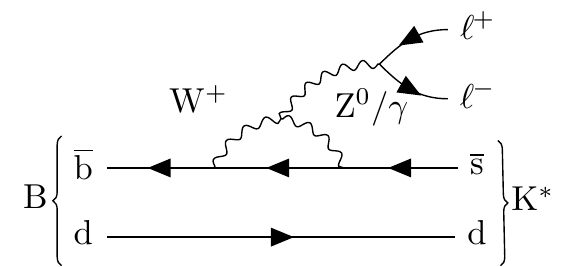}
	\includegraphics{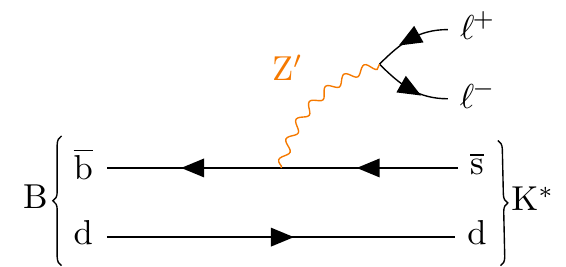}
	\caption{Leading order SM process for the decay $\BKstarLL$ (left) and one possible BSM realization of the process with a $Z'$ boson (right).}
	\label{app:feynman:RKstarLL}
\end{figure}

The decays $\BKstarLL$ have been studied intensively and the experimental results suggest that the underlying $\Pbottom \rightarrow \Pstrange \Plepton \Plepton$ transition could be affected by BSM physics in the final state containing muons\cite{rkstar-1,rkstar-2,rkstar-3,rkstar-4,rkstar-5}.
The ratio of branching ratios
\begin{equation}
	R_{\PKstar} = \frac{\mathcal{B}(\PB \rightarrow \PKstar \Pmu \Pmu)}{\mathcal{B}(\PB \rightarrow \PKstar \Pe \Pe)} \approx 1
\end{equation}
is well suited for testing lepton flavor universality\cite{rkstar-6}. In the SM the ratio is predicted to be close to unity, i.e.\ the coupling of the gauge boson does not distinguish between electrons and muons. The theoretical uncertainties originating from the form factors of the hadronic transition cancel in the ratio, making it an excellent probe for new physics.\cite{rkstar-7} In Figure~\ref{app:feynman:RKstarLL} the leading order SM process is shown together with a possible new physics model involving a $Z'$ boson.

\subsection{Results}
\begin{figure}
	\centering
	\includegraphics[width=0.45\linewidth]{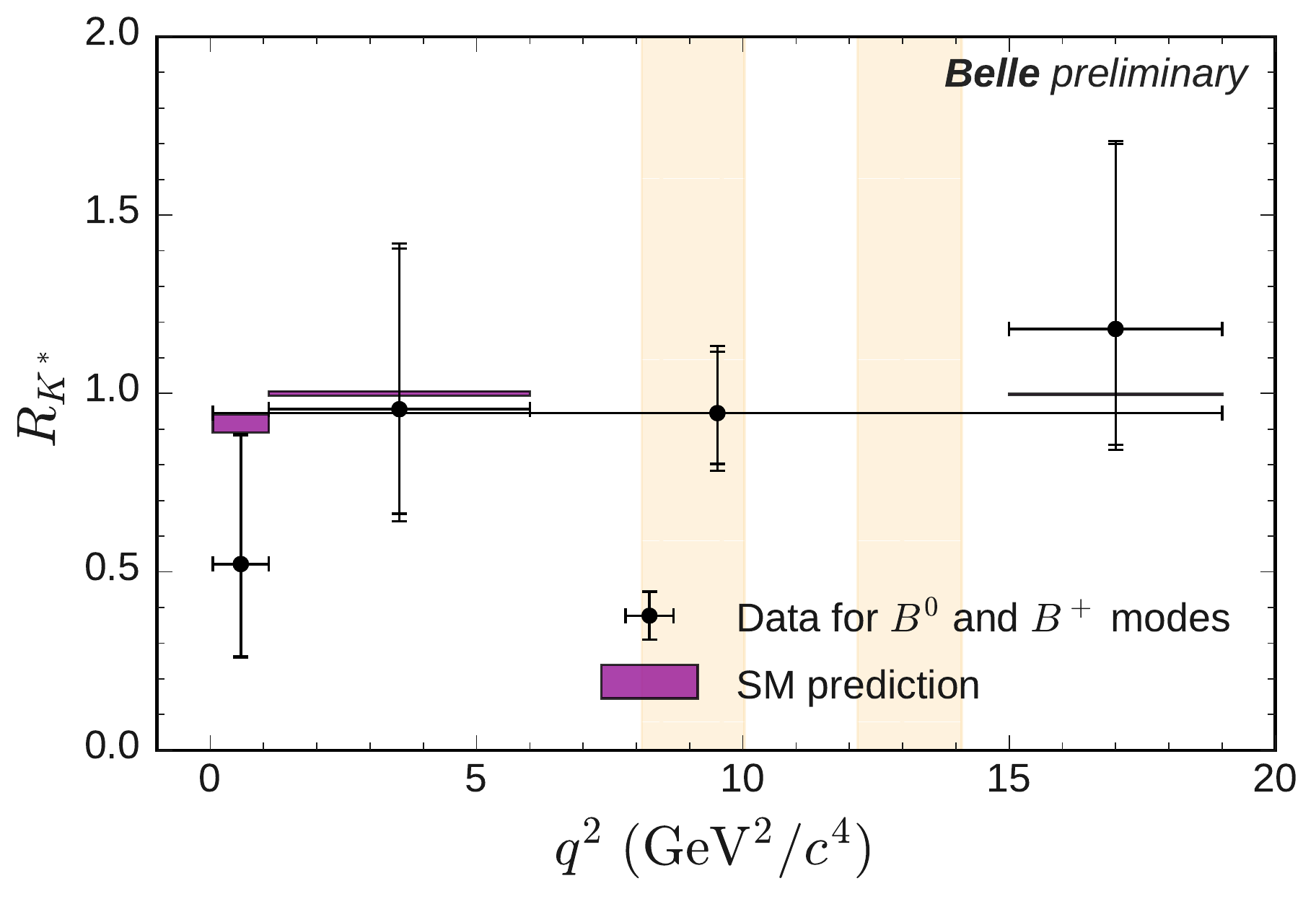}
	\caption{Results for $R_{\PKstar}$ compared to the SM predictions. The separate vertical error bars indicate the statistical and total uncertainty.}
	\label{fig:BKstarLL:ratio}
\end{figure}

The key improvements of the presented analysis are an improved reconstruction of the decay with a hierarchical neural network approach and the optimization of the analysis to measure the ratio and not the individual branching ratios. Additionally, the presented analysis includes for the first time the final state with a $\PKstar^+$ meson, i.e.\ the measurement of the ratio $R_{\PKstar^+}$. For a detailed description of the analysis see Reference~\cite{wehle}.
The measured ratio of $R_{\PKstar}$ over the whole $q^2$ range is 
\begin{equation}
	R_{\PKstar} = 0.94^{+0.17}_{-0.14}\pm 0.08\ .
\end{equation}
The value of $R_{\PKstar}$ in individual bins of $q^2$ is shown in Figure~\ref{fig:BKstarLL:ratio} and tabulated in Table~\ref{tab:BKstarLL:ratio}.
 
\begin{table}
	\centering
	\caption{Result for $R_{\PKstar}$ , $R_{\PKstar^0}$ and $R_{\PKstar^+}$. The first uncertainty is statistical and the second is systematic.}
	\label{tab:BKstarLL:ratio}
	\begin{tabular}{lccc}
		\hline
		\hline
		$q^2$  in GeV$^2$/c$^4$ & All modes & $B^0$ modes & $B^+$ modes  \\
		\hline
		$[0.045, 1.1]$ & $ 0.52^{+0.36}_{-0.26}\pm0.05 $ & $ 0.46^{+0.55}_{-0.27}\pm0.07 $  & $ 0.62^{+0.60}_{-0.36}\pm0.10 $  \\
		$[1.1, 6]$  & $ 0.96^{+0.45}_{-0.29}\pm0.11 $  & $ 1.06^{+0.63}_{-0.38}\pm0.13 $  & $ 0.72^{+0.99}_{-0.44}\pm0.18 $  \\
		$[0.1, 8]$  & $ 0.90^{+0.27}_{-0.21}\pm0.10 $  & $ 0.86^{+0.33}_{-0.24}\pm0.08 $  & $ 0.96^{+0.56}_{-0.35}\pm0.14 $  \\
		$[15, 19]$  & $ 1.18^{+0.52}_{-0.32}\pm0.10 $  & $ 1.12^{+0.61}_{-0.36}\pm0.10 $  & $ 1.40^{+1.99}_{-0.68}\pm0.11 $  \\
		$[0.045, ]$ & $ 0.94^{+0.17}_{-0.14}\pm0.08 $  & $ 1.12^{+0.27}_{-0.21}\pm0.09 $  & $ 0.70^{+0.24}_{-0.19}\pm0.07 $  \\
		\hline
		\hline
	\end{tabular}	
\end{table}

\end{document}